\documentclass[reprint,onecolumn,amsmath,amssymb,notitlepage]{revtex4-1}

\usepackage{graphicx,color}
\usepackage{hyperref}

\makeatletter
\hypersetup{
	    pdfcreator={Cesar Henrique Comin},
		colorlinks=true,       		% false: boxed links; true: colored links
    	linkcolor=black,          	% color of internal links
    	citecolor=black,        		% color of links to bibliography
    	filecolor=black,      		% color of file links
		urlcolor=black,
		bookmarksdepth=4
}
\makeatother
\begin{document}

\title{The relationship between structure and function in locally observed complex networks}

\author{Cesar H. Comin$^1$}
\email{Email: chcomin@gmail.com}
\author{Matheus P. Viana$^1$}
\author{Luciano da F. Costa$^1$}

\affiliation{$^1$S\~ao Carlos Institute of Physics, University of S\~ao Paulo, S\~ao Carlos, S\~ao Paulo, Brazil}

\begin{abstract}

Recently, some studies started to unveil the wealthy of interactions that occur between groups of nodes when looking at the small scale of interactions taking place in complex networks. Such findings claim for a new systematic methodology to quantify, at node level, how a dynamics is being influenced (or differentiated) by the structure of the underlying system. Here we define a new measure that, based on dynamical characteristics obtained for a large set of initial conditions, compares the dynamical behavior of the nodes present in the system. Through this measure we find that the geographic and Barab\'asi-Albert models have high capacity for generating networks that exhibit groups of nodes with distinct dynamics compared to the rest of the network. The application of our methodology is illustrated with respect of two real systems. In the first we use the neuronal network of the nematode \emph{Caenorhabditis elegans} to show that the interneurons of the ventral cord of the nematode presents a very large dynamical differentiation when compared to the rest of the network. The second application concerns the SIS epidemic model on an airport network, where we quantify how different the distribution of infection times of high and low degree nodes can be, when compared to the expected value for the network.
\end{abstract}

\maketitle

\section{Introduction}

Given that complex systems are almost invariantly composed by a large number of interacting elements, they can be efficiently represented and studied in terms of complex networks~\cite{newman:2003,barabasi:2002,costa:2011}. In this representation, their structural and dynamical properties can be extracted and investigated. Typically, the structure of such networks is quantified in terms of several measures~\cite{costa:2007}, reflecting different properties of the respective topology (e.g. node degree, shortest paths,
centralities) and geometry (e.g. arc length distances, angles, spatial density). 

A great deal of the investigations about structure and function in complex systems has focused on trying to predict the dynamics from specific structural features~\cite{vespignani:2008,boccaletti:2006,dorogovtsev:2008}. Such an ability would provide the means for effectively controlling real-world systems \cite{liu:2011}. Despite the growing number of studies devoted to this problem, the knowledge about the relationship between the structural and dynamical properties remains incipient because of three main reasons: (a) dynamics is often summarized in terms of global statistics, which overlooks its intricacies; (b) the investigation often focuses on linear relationships such as correlations between structural and dynamical features; and (c) several effects, such as initial conditions, network topology, stochasticity or dynamical differences from node to node are not selectively fixed or controlled. Still, there are some notable examples of local analysis previously done on complex networks. Garde\~nes et al. \cite{gardenes:2007} studied how synchronization takes place on heterogeneous random systems, in comparison to their homogeneous counterparts. They found that the systems differ by the way smaller synchronized groups are formed while increasing the coupling strength of the Kuramoto oscillators. Kitsak et al. \cite{kitsak:2010} used the k-shell index to find nodes having a high potential to spread a disease in a network. Their fundamental result was that it is possible to predict the number of infected nodes of the entire system by knowing only the k-shell of the initial infected node. Still on the subject of epidemics, there are other works concerning local analysis \cite{barthelemy:2004,satorras:2002} as well as the use of temporal networks to show the importance of initial conditions \cite{rocha:2011}. Another good example of local structural analysis that can be translated to dynamical behavior is the study of motifs \cite{milo:2002,sporns:2004}.

In this article we propose a novel methodology to quantify how much the dynamics at each node differentiates from the dynamics at the other nodes as a consequence of specific aspects (e.g. local anisotropies) of the network structure. This is accomplished by simulating the investigated dynamics for a large number of initial conditions and checking how much a given property of the dynamics at a node (e.g. entropy of the time series at that node) deviates from the overall dynamics. The level at which a node $i$ ``feels'' the structure differently from the other nodes is quantified in terms of a parameter $\alpha_i$.  In this way, the proposed methodology addresses the three shortcomings mentioned above by: (i) being local, i.e. it is applied for each individual node; (ii) by not imposing any specific kind of relationship between dynamical and structural features; and (iii) isolating each (above mentioned) condition that can affect the dynamics.

Several important findings have been obtained by using this methodology. Our results show that the nodes feel rather distinctly the structure in most of the considered situations. While the Erd\H{o}s-R\'enyi model \cite{erdos:1960} does not show any dynamical differentiation, the geographic model of Waxman \cite{waxman:1988} presents fluctuations that naturally originates different dynamical groups through the density of connections. The Barab\'asi-Albert model \cite{barabasi:1999} shows a rather distinct behavior for the highly connected nodes of the network, which end up having a very distinct dynamics related to the rest of the network, being even more extreme than the topological differences. When considering a real network of the nematode \emph{Caenorhabditis elegans}, we find that there exists a group of neurons where the local topology influences rather distinctly on the spike rate of the signals. Finally, the study concerning epidemic dynamics shows that the first infection time of a node can have different levels of variability depending on the degree of the node.

\section{Methodology}

\subsection{Measuring the differentiation}

In order to apply our methodology we begin with a network having $N$ nodes, which will be fixed through the entire process, and execute $M$ times a dynamics on it. Each execution starts with a randomly sorted initial condition. It is important to note that depending on the dynamics being studied we can have a specific set of initial conditions that take the system to a particular state, so this state will rarely be accessed by sampling. That is not a problem in our method because we are analysing the dynamics for the set of initial conditions imposed, that is, we are studying the signals that are in fact observed.

\begin{figure}[!htbp]
  \begin{center}
    \includegraphics[width=0.5\linewidth]{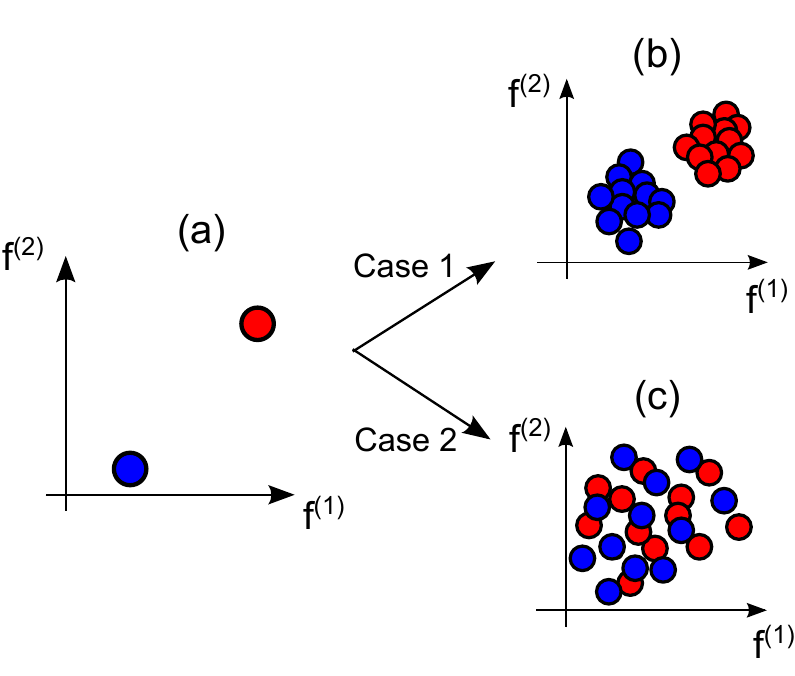} \\
    \caption{Example of topological diferentiation for two nodes. Projecting the signal characteristics for a single realization of the dynamics (a) is not suficient to discern the topological influence. We need to consider many distinct initial conditions in order to infer if the topology diferentiates (b) or not (c) the dynamics of the nodes.}~\label{f:real}
  \end{center}
\end{figure}

In possession of the dynamic signals for the $M$ realizations, we need a mechanism to represent them in order to compare their behavior. In our case we use dynamical measures that tries to extract the most relevant information about the signals. Let $F$ be one of such measures, after the many realizations we get a set of observed values {$f_{i,r}$}, where $i$ is the node index and $r$ the realization. These values can vary through four distinct mechanisms: (a) initial condition, (b) network topology, (c) particular dynamics and (d) stochasticity. Our objective here is to study only the relation between the initial condition and the topology, and so we use dynamics having identical equations for every node and consider stochastic variations to be small. With these restrictions the only variations we can observe on the values of $F$ are: (a) fluctuations on the dynamical values of a given node, which given the fact that the topology is static, \emph{can only be caused by the variation of the initial condition}. (b) Differences on the mean values of $F$ for distinct nodes, that because of the properties assumed \emph{can only be caused by the topological differences of the nodes}. The term \emph{topological difference} need to be used with care, because unless in very specific cases where the network is perfectly symmetric (e.g., a lattice with toroidal boundary), the topology of two given nodes is rarely identical, that is, we can always find a structural characterization that will have distinct values for them. Nevertheless, since the networks we use are not regular, every significant dynamical difference we observe must be caused by the topology. It is also important to note that the reverse is not true, if the dynamics of two nodes appear to be the same, their topologies are still distinct, what happened is that both nodes \emph{felt} the topology in the same manner. An example of this last case is the diffusion dynamics on graphs \cite{lovasz:1993}, in which the equilibrium behavior depends only on the degree of the nodes, that is, although the nodes possess distinct general topology, the localized characteristic of the dynamics allows only the degree to differentiate the nodes. The two cases we may come across are shown in Figure \ref{f:real}.

In order to quantify the difference of the values obtained for each node, we use a statistic test that will be defined on the next section. Through this statistic we can identify if the difference of the means of the dynamical values obtained are in fact significant, that is, we are quantifying the difference between the dynamics of the nodes normalized by the intrinsic fluctuations caused by the initial condition. The distance between every pair of nodes is them represented by the matrix $\Xi$, where each line $i$ and column $j$ represents the distance obtained between the nodes $i$ and $j$, that is,

    \begin{equation}
        \Xi_{ij}=d_h(i,j)\label{eq:elem_dist}
    \end{equation}

where $d_h(i,j)$ is the distance defined by equation \ref{eq:dist_corr}. Finally, we can define our mean dynamical differentiation measure, $\alpha$, as the mean values of each line of this matrix

    \begin{equation}
        \alpha_{i}=\frac{1}{N-1}\overset{N}{\underset{j=1}{\sum}}\Xi_{ij}\label{eq:alpha}
    \end{equation}

    The standard procedure now would be to calculate the statistical significance of the observed values of $\alpha$, but since we are concerned with the comparison between the distances, and not on their absolute values, this does not need to be performed. To carry out the comparison, we construct an histogram of the obtained values of $\alpha$. Having in mind that $\alpha$ is relative to some dynamical characteristic, we can have distinct histograms relative to the desired characterization.

\begin{figure}[!htbp]
  \begin{center}
  \includegraphics[width=0.5\linewidth]{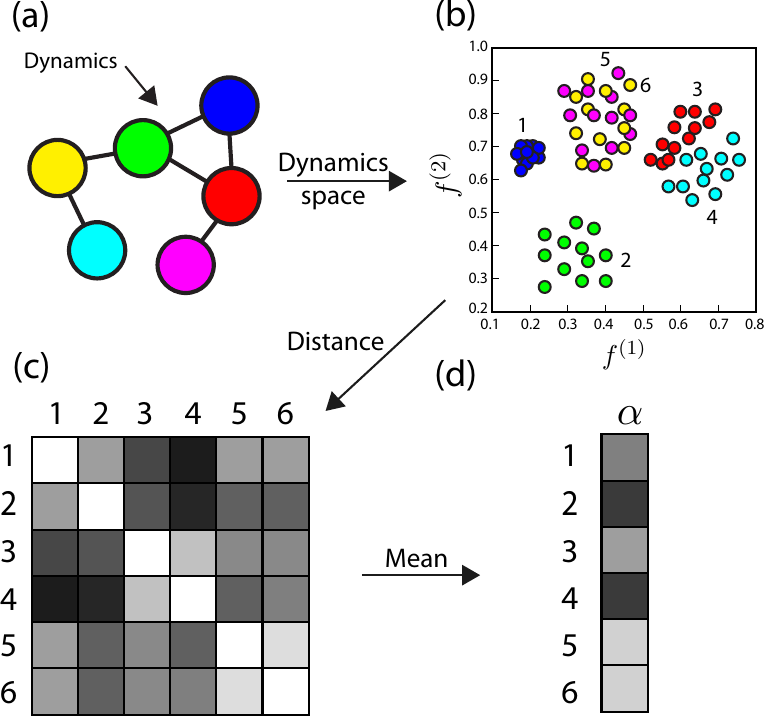} \\
  \caption{Example application of the methodology. (a) We execute 12 runs of the dynamics with different initial conditions and (b) project the obtained signals on the measure space, which in this case is 2-dimensional. (c) The matrix $\Xi_{ij}$ is obtained using equation \ref{eq:dist_corr}, and (d) its mean is taken in order to obtain the $\alpha$ of each node.}~\label{f:modelo2}
  \end{center}
\end{figure}

    What we search for are particular behaviors of the histograms, for example, it is expected that a single node with very distinct dynamics compared to the rest of the network will have a very large $\alpha$ value. It is important to observe that although we presented the methodology for a single measure, nothing prevents us from calculating the distances using simultaneously various dynamical measures. In Figure \ref{f:modelo2} we show an example application of the presented model.

\subsection{Distance measure}

\begin{figure}[!htbp]
  \vspace{0.3cm}
  \begin{center}
  \includegraphics[width=0.9\linewidth]{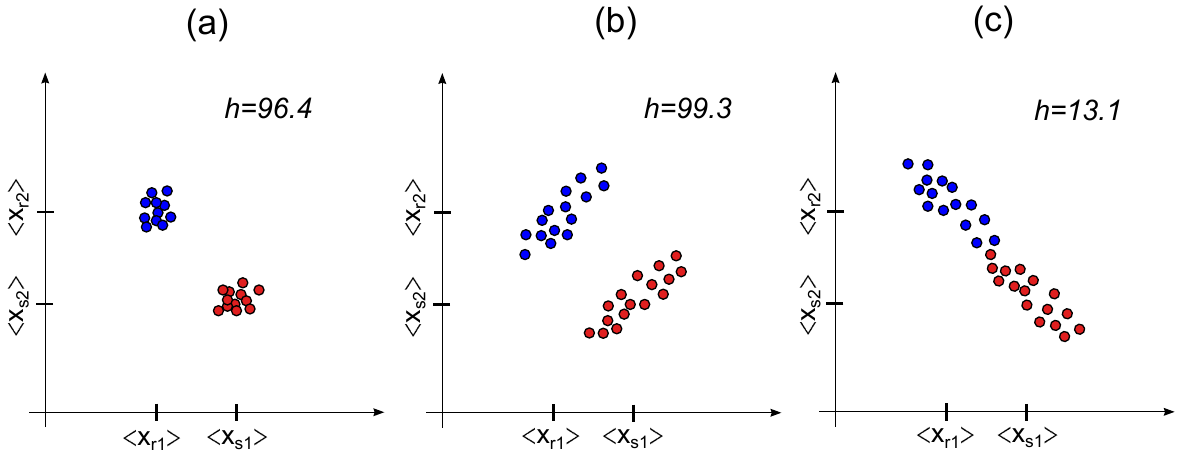} \\
  \vspace{0.5cm}
  \caption{Example of Hotelling distances. The three Figures present the same Euclidean distance between the mean point of the two groups, but in cases (a) and (b) the statistical distance is more significant than in (c). The respective Hotelling distances are shown in each Figure.}
  ~\label{f:hotelling} \end{center}
\end{figure}

In this section we present and motivate the statistical test that will be used throughout the paper. Suppose that we have a set of points inserted in a $m$-dimensional space and these points form two distinct groups (see Figure \ref{f:hotelling}). An immediate way to quantify the separation between the groups is by using the Euclidean distance between the center of mass of each group, given by

\begin{equation}
	d(r,s)=\sqrt{\overset{m}{\underset{i}{\sum}}(\langle x_{ri}\rangle-\langle x_{si}\rangle)^{2}},
	\label{eq:eucl}
\end{equation}

where $r$ and $s$ are the indices of the groups and $\langle x_{ri}\rangle$ represents the mean (or center of mass) of the points in group $r$ on the $i$-th dimension. The disadvantage of the Euclidean distance is that it does not take into account the dispersions of the groups from which the distance is being measured. Following Figure \ref{f:hotelling}, if we have two random variables $X_r$ and $X_s$ with the realized values $\{x_r\}$ and $\{x_s\}$, marked respectively in blue and red in the figure, clearly the two cases shown in Figures \ref{f:hotelling}(a) and (b) have a more significant distance between their mean than the case in Figure \ref{f:hotelling}(c), although the Euclidean distance is the same. There are many ways to take into account the dispersions of the groups, here we use the Hotelling statistic \cite{hotelling:1931}, which considers the variance of each group in the direction defined by the line that passes between the two means. The distance, $h$, between two groups is defined by

\begin{equation}
	h^{2}=\frac{n_{r}n_{s}}{n_{r}+n_{s}}(\langle\vec{x}_{r}\rangle-\langle\vec{x}_{s}\rangle)'\Sigma_{m}^{-1}(\langle\vec{x}_{r}\rangle-\langle\vec{x}_{s}\rangle),
        \label{eq:hot}
\end{equation}

where $\langle\vec{x}_{r}\rangle$ is the average position of group $r$. The variable $\Sigma_m$ is the estimation of the equivalent covariance matrix of each group, given by

\begin{equation}
	\Sigma_{m}=\frac{n_{r}\Sigma_{r}+n_{s}\Sigma_{s}}{n_{r}+n_{s}-2},
\end{equation}

where, $\Sigma_r$ and $\Sigma_s$ are the covariance matrix of the groups. In Figure \ref{f:hotelling} we show the values of $h$ for each case.

The distance we considered is an hybrid version of both Euclidean and Hotelling metrics. This is required in order to avoid singularities for set of nodes with zero variance. Since we are not interested on the absolute value of the distance, but on the comparison of values between groups, we can define a new hybrid distance, given by

     \begin{equation}
        d_{h}(r,s)=\frac{h}{h+1}d(r,s)\label{eq:dist_corr}
     \end{equation}

where $d(r,s)$ is the usual Euclidean distance between the center of mass of groups $r$ and $s$.

Finally, in cases we have to calculate the distances through more than one variable, it is necessary to normalize them so as to give a fair comparison. In order to do so we calculate the standard score of each measure $x$, given by

     \begin{equation}
        \hat{x}=\frac{x-\langle x\rangle}{std(x)}
     \end{equation}

where $std(x)$ is the standard deviation of $x$. Since through the paper we always use the standardized version of the values, we simplify the notation by calling $\hat{x}$ just by $x$.

\section{Studied systems}

In this section we briefly present the topologies and dynamics where our methodology was applied.

\subsection{Network models}
% O revisor pediu que explicasse os modelos, mas nao queria colocar uma secao so para modelos.

In our study we use three network models to compare the methodology applied to three distinct scenarios. The first is the traditional Erd\H{o}s-R\'enyi (ER) \cite{erdos:1960} model that connects every possible pair of nodes with a probability $p$, originating a Poisson degree distribution representing a completely random graph.

The second scenario is when every connection has a cost associated with it, usually represented by a geographic network where the nodes have a spatial position. A commonly used mechanism to model this behavior is due to Waxman \cite{waxman:1988}, who randomly placed the nodes in a [1,1] grid and, for every pair $(i,j)$ of nodes, defined a probability of this pair being connected, given by

\begin{equation}
    P(i,j)=\beta e^{-d(i,j)/d_{0}}\label{eq:waxman}
\end{equation}

where $d(i,j)$ is the Euclidean distance between nodes $i$ and $j$, $\beta$ tunes the density of edges, and $d_0$ sets the typical size of the connections. This generates a topology where a long range shortcut is rare to occur. Because of this, one of the main characteristics of this model is that, for the most commonly used values of $\beta$ and $d_0$, it does not exhibit the small world behavior.

Another class of networks, usually called power-law, is greatly represented by the Barab\'asi-Albert model \cite{barabasi:1999}, where two principles, namely growth and referential attachment are used to model the usual power-law degree distribution ($P_{k}\approx k^{-3}$) that is observed in many real systems \cite{clauset:2009}. Because of the power-law distribution, the network generated shows high degree fluctuations, which originates a small portion of nodes that possesses a very large degree when compared to the rest of the nodes.

\subsection{Dynamics used}
{\bf Integrate-and-fire dynamics}

%Since we are not interested in studying the dynamics \emph{per se}, but use it as a tool to present our methodology, we choose a further simplified discrete integrate-and-fire dynamics given by
%
%\begin{equation}
%    V_{i}(t+1)=
%    \begin{cases}
%    V_{i}(t)+\underset{j=1}{\overset{N}{\sum}}\underset{f}{\sum}A_{ij}\delta(t-t_{j}^{f}) & \text{if} \;\; V_{i}(t)<T_{l}\\
%    \underset{j=1}{\overset{N}{\sum}}\underset{f}{\sum}A_{ij}\delta(t-t_{j}^{f}) & \text{if} \;\; V_{i}(t)\geq T_{l},
%    \end{cases}
%    \label{eq:intfire_simpl}
%\end{equation}
%
%where $V_i(t)$ is the membrane potential of neuron $i$ at time $t$, $t_j^f$ the instant of the $f$-th spike of neuron $j$, $\delta(x)=1$ when $x=0$ and $\delta(x)=0$ otherwise, and $\sigma$ the couple strength. When $V_i$ reaches the threshold $T_l$, the neuron fires a unitary signal to all its neighbors and $V_i$ is reseted to zero. The values for membrane potential at $t=0$ were randomly sorted with uniform probability inside the range $[0,T_l]$.

Our first application of the method will be related to the transmission of neuronal signals, modeled by the integrate-and-fire dynamics \cite{lapicque:1907,brunel:2007}. This model treats the neuron as an integrator with a hard threshold limit, $\mathcal{T}$. The actions of a given neuron $i$ along the time is stored in the binary time series $s_i(t)$, which indicates that the neuron spikes at instant $t$ whether $s_i(t)=1$. Using this time series we define the spike rate of a neuron as

\begin{equation}
	r_{i}=\frac{1}{T_{sim}-T_{est}}\underset{t=T_{est}}{\overset{T_{sim}}{\sum}}s_i(t),
	\label{eq:taxa}
\end{equation}

where $T_{sim}$ is the total simulation (or experiment) time and $T_{est}$ is a long enough time, found empirically trough   prior simulations, so as that $r_i$ do not significantly change after $T_{est}$ (we say that the dynamics stabilized). This measure corresponds to the average number of spikes during the considered interval, and it is widely used in neuroscience, since many neurons codify the stimulus amplitude through the rate of spikes~\cite{rieke:1999}. 

\phantom{aa}

\noindent{\bf SIS model:}

In the SIS model each node can be in one of two states: infected or susceptible. The spread of the disease between neighbors happens with a rate $\beta$, interpreted as the probability per time step that the disease will spread from an infected node to a susceptible one. Each node returns to the susceptible state with a rate $\gamma$ which, without loss of generality, we define as being $\gamma=1$. There are many ways to simulate the start of this disease on a network. Here we chose to randomly select with equal probability a single node and turn it to the infected state. After iterating the dynamics for a sufficiently long time we keep the simulation data if the disease has spread to the entire network, otherwise we start another simulation. The measure we use in this case is the first infection time, $I_f$, defined as the iteration where the node got its first infection. The value of $I_f$ for a node will strongly depend on the initial condition, so it will be a good case study for our method.

\section{Results}

\subsection{Integrate-and-fire on random network models}

\begin{figure}[!htbp]
      \begin{center}
      \includegraphics[width=0.8\columnwidth]{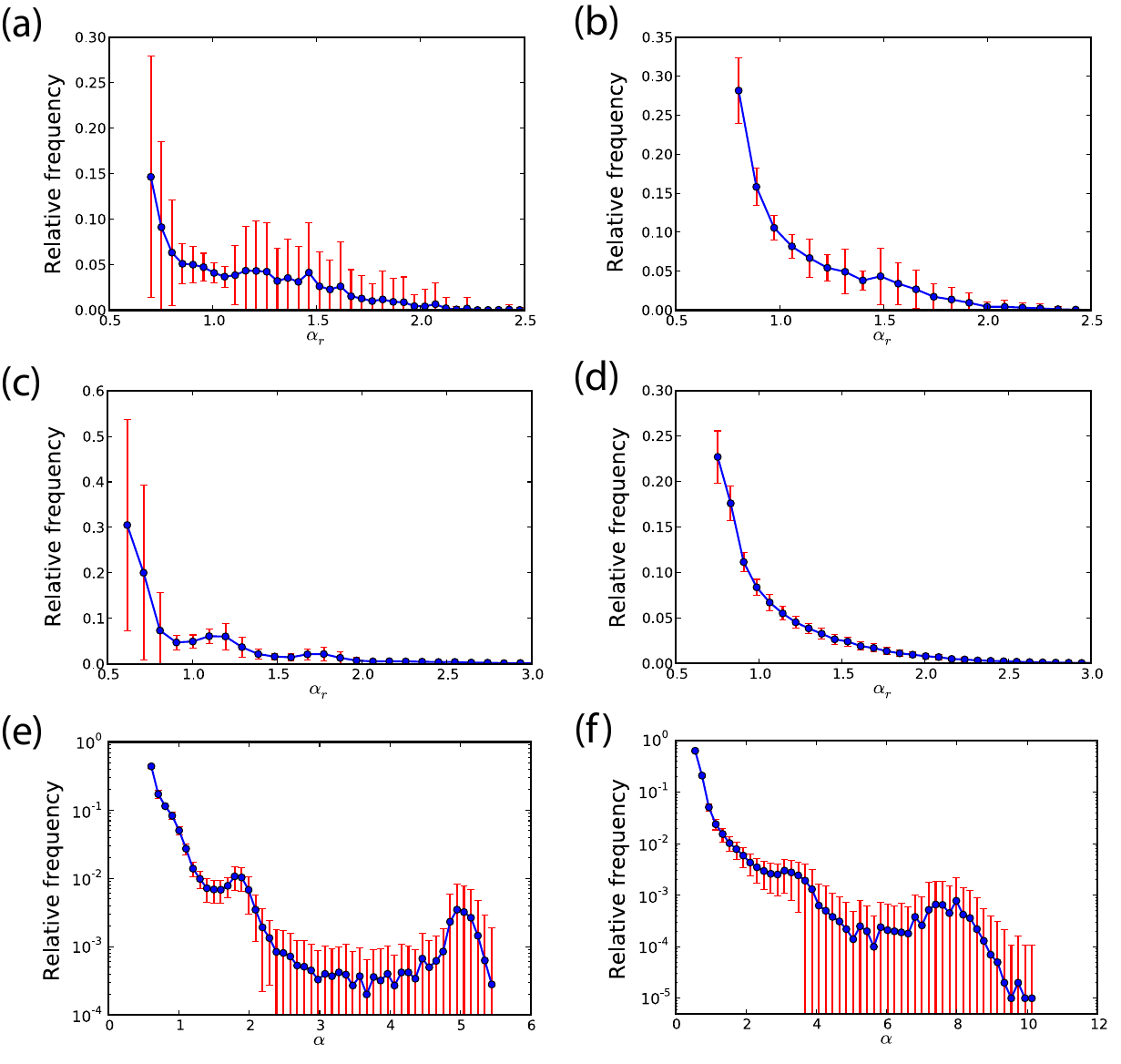} \\
      \caption{$\alpha$ values obtained for the integrate-and-fire dynamics on random networks. For each case we generated 100 networks and obtained the histogram of $\alpha$ for each network. The plots show the mean and standard deviation of these histograms. The networks were generated using (a) and (b) the geographic model, (c) and (d) the ER model, (e) and (f) the BA model. Graphics on the left were obtained using $\mathcal{T}=8$, and the ones on right with $\mathcal{T}=10$.}~\label{f:alfa_modelos} \end{center}
    \end{figure}

We compare the differentiation relative to the spike rate feature, which we call $\alpha_r$ for different network topologies, namely Erd\H{o}s-R\'enyi (ER) \cite{erdos:1960}, Barab\'asi-Albert (BA) \cite{barabasi:1999} and Waxman geographic model \cite{waxman:1988}. In Figure \ref{f:alfa_modelos}(a) we show the result obtained for the geographic model with $N=1000$ and $\langle k\rangle=10$ and an integrate-and-fire dynamics with $\mathcal{T}=8$ taking place on the network. To obtain statistical significance we use 100 different generated networks, and each network is subjected to $M=1000$ realizations of the dynamics with different initial conditions. We construct histograms of $\alpha_r$ obtained for each generated network and show in Figure \ref{f:alfa_modelos}(a) the mean value and standard deviation the set presents. We see that the frequency of nodes with small $\alpha_r$ has a large variation, which is caused by the intrinsic fluctuations in the dynamics of the many nodes with similar spike rate present on the network. It is feasible to think that this fluctuation would decay as $\alpha_r$ increases, but this is true only for intermediate $\alpha_r$, while at high values of $\alpha_r$ we observe a sudden increase of fluctuation. Additionally, the mean value stays at an almost constant value for $\alpha_r$ in the range $[0.9,1.5]$. This is caused by the high potential of the geographic model to display structure fluctuations, originating regions with higher density when compared to the rest of the network. These regions alter significantly the spike rate of the nodes, and high differentiated groups appears. In Figure \ref{f:alfa_modelos}(b) we apply the same procedure to calculate $\alpha_r$ in the geographic model, only changing the dynamic threshold to $\mathcal{T}=10$. It is clear that the groups are no longer distinguishable. This is so because the threshold is now so large that even the topological fluctuations cannot differentiate a significant number of nodes, when compared to the ones with small $\alpha_r$.

We apply the same procedure used for the geographic model to the ER networks with $N=1000$ and $\langle k\rangle=10$. In Figure \ref{f:alfa_modelos}(c) we show the information about the obtained histograms for $\mathcal{T}=8$, which makes clear that this model exhibits much smaller fluctuations. This is caused by the much shorter geodesic distances that the model exhibits, when compared to the geographic counterpart, which creates a more compact network. We also show in Figure \ref{f:alfa_modelos}(d) the case $\mathcal{T}=10$ for the ER model, where we see a decaying behavior expected for a random Poissonian system.

The third investigated model is the BA with $N=1000$ and $\langle k\rangle=6$. Figures \ref{f:alfa_modelos}(e) and (f) show the log-scale histogram for, respectively, $\mathcal{T}=8$ and $\mathcal{T}=10$. In both cases we observe a significantly high peak for large $\alpha_r$, which we found to be related to the network hubs (nodes with very high degree). This result was expected, given our observations of large fluctuations on the geographic network, but this is not the main result for the BA model. The important result is that although the power-law degree distribution of the model has a continuous decaying behavior, the histogram of $\alpha_r$ shows small values for intermediate $\alpha_r$ and increases for large $\alpha_r$. This behavior can be interpreted as follows: the dynamical differentiation of a hub is, as expected, very large, but a node with almost equal degree can end up with a much smaller differentiation, having dynamics more similar to the low degree nodes. This result confirms the important role that hubs have in complex systems, not only in the sense of being central, but also in having a different purpose to the network dynamics.

\subsection{Integrate-and-fire on the network of the Caenorhabditis elegans}

Although many interesting properties arise when studying random network models, it is on real networks that the dynamical differentiation analysis can show its real potential. To show this we now apply the methodology to the \emph{C. elegans} neuronal network. In this network, each node represents a neuron and two nodes are connected if there exists some kind of directed communication between them (e.g. synapses, gap junctions, etc). The data was compiled by Chen et al. \cite{chen:2006,varshney:2011} and obtained from \cite{base_elegans}. The network has 279 nodes and $\langle k \rangle = 22.4$.

Although we motivate the method with a real network, it is important to note that our dynamics does not take into account many signal particularities that arise for real neurons \cite{Izhikevich:2004}, therefore we are looking for a coarse grained description of the neurons inside the network. We will show that, even with this simplified description, it is still possible to observe some interesting phenomena.

\begin{figure}[!htbp]

  \begin{center}
      \includegraphics[width=0.4\linewidth]{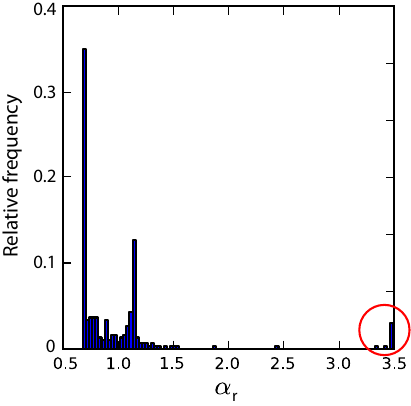}
      \caption{Histogram of the $\alpha$ values obtained for the integrate-and-fire signals taking place on the \emph{C. elegans} network. 1000 realizations of the dynamics were made with different initial conditions. The $\alpha$ was calculated using the spike rate  measure.}~\label{f:alpha_uni_elegans}
  \end{center}
\end{figure}

\begin{figure}[!htbp]
  \begin{center}
      \includegraphics[width=0.5\linewidth]{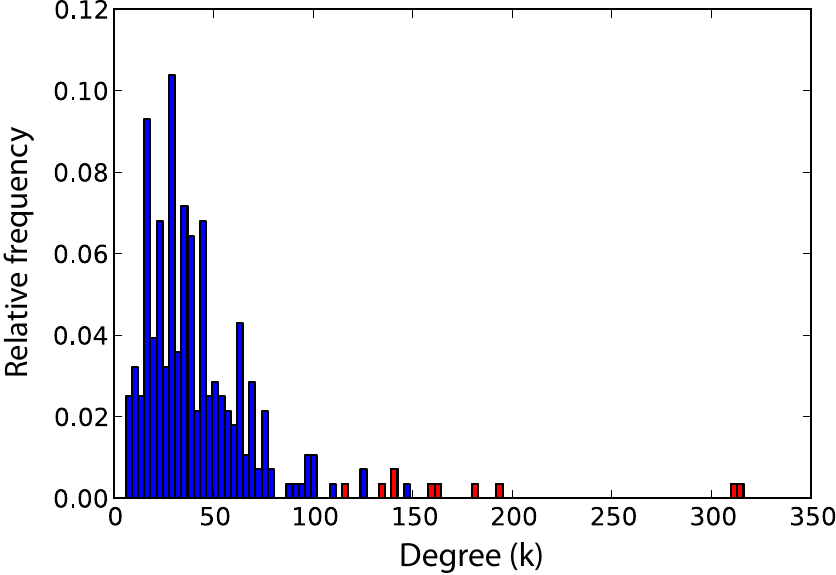}
      \caption{Degree distribution of the \emph{C. elegans} network. Red bars indicate the nodes present in Table \ref{tab:c_elegans}, i.e., the nodes indicated in Figure \ref{f:alpha_uni_elegans} as having the highest $\alpha$ values.}~\label{f:degree_elegans}
  \end{center}
\end{figure}

We begin by showing in Figure \ref{f:alpha_uni_elegans} the histogram of $\alpha$ relative to the spike rate, $\alpha_r$. An immediate result we can observe is that there is a group with high differentiation, indicated with a red circle, which is somewhat similar to that observed for the BA random model. With this in mind we plot in Figure \ref{f:degree_elegans} the degree histogram of the network, indicating in red the nodes inside the observed group. We see that the nodes with high differentiation possess a high degree in the network, but there are some high degree nodes that do not show a distinct dynamics. It is clear that the topology influence does not occur merely by the degree of the nodes. There is a particular relation between these high differentiated nodes that make their dynamics very peculiar when compared to the rest of the network. The nodes inside the red circle in Figure \ref{f:alpha_uni_elegans} are known as the interneurons of the ventral cord of the \emph{C. elegans}. They are well recognized for possessing a high number of synapses \cite{white:1986}, given that they make the bridge between sensory and motor neurons without much restriction on the type of transmitted signals (some classes of interneurons are known for receiving only a specific type of signals).

In table \ref{tab:c_elegans} we present the traditional names of these neurons and some information about their spatial and topological distance. Position refers to the spacial localization of each neuron relative to the axis that goes from the head (value 0) to the tail (value 1) of the nematode. We see that the majority of the described neurons are on the head (more specifically, in the nerve ring of the nematode \cite{ware:2004,wood:1988}), with the exception of PVCL and PVCR that are on the tail. In the table, D1 is the mean topological distance between these neurons, that is, given a node $i$ we measure how many edges we need to travel in order to go to node $j$, and we take the mean of this distance for all $j$ inside the differentiated group. $D1=1$ means that the neuron is a neighbor, or receives a direct signal, of all the other neurons shown in the table. The feature D2 complements D1 as it shows the topological distance between the given neuron and all other neurons, excluding those present in the table. We see that in all cases the differentiated nodes are closer between themselves than with the rest of the network, an effect partially provoked by their high degree. That is, besides having a high degree, these nodes are well connected between themselves, originating a high capacity of communication inside the group, and rendering their dynamics very distinct in comparison to the rest of the network. In order to illustrate these neurons, we show in Figure S1 of the supplementary material the position of each neuron inside the nematode.

\begin{table}
\caption{Calculated topological values for the interneurons of the ventral cord of \emph{C. elegans}. See text for explanation about D1 and D2.}\label{tab:c_elegans}

\begin{centering}
{\small }\begin{tabular}{c c c c c c c c c c c}
\hline
{\small Neuron} & {\small AVAL} & {\small AVAR} & {\small AVBL} & {\small AVBR} & {\small AVDL} & {\small AVDR} & {\small AVEL} & {\small AVER} & {\small PVCL} & {\small PVCR}\tabularnewline
\hline

{\small Position} & 0.13 & 0.13 & 0.15 & 0.15 & 0.16 & 0.16 & 0.13 & 0.14 & 0.82 & 0.82\tabularnewline

{\small D1} & 1.00 & 1.00 & 1.00 & 1.44 & 1.33 & 1.22 & 1.33 & 1.22 & 1.00 & 1.00\tabularnewline

{\small D2} & 1.76 & 1.77 & 1.81 & 1.80 & 1.97 & 1.87 & 1.90 & 2.09 & 2.04 & 2.03\tabularnewline
\hline
\end{tabular}
\par\end{centering}{\small \par}

\end{table}

\subsection{Epidemics on the airport network}

In order to show that our approach can be applied to a completely different system, we also study the first infection time, $I_f$, of nodes going through an epidemics dynamics. The network we used is the world-wide airport network, which describes the flying routes between a large number of airports through the world. Each node is an airport (3302 in total) and an edge indicates that there is a flight between two airports. The data was obtained from \cite{airport}. 

The value of $I_f$ for a node strongly depends on its relative position with respect to where the epidemics has started, so for each initial condition the node will have many different $I_f$ values. Still, the local topology of the node can influence the mean time it takes for the infection to arrive. Suppose we take two nodes with similar local topologies. It is expected that their $I_f$ will be slightly different, but how different they really are? If the topologies of these nodes completely define their dynamics, i.e., $I_f$ does not vary for the different initial conditions, them the difference in their dynamics, caused by the topology, is significant. On the contrary, if their mean values are close but the topology does not have a strong influence, we have that with different initial conditions it is possible to reach many different dynamical states. This means that the nodes are not being significantly differentiated by their local topology. We reinforce that this is different from simply comparing their mean values, as we are taking into account both the variation caused by the initial condition and the network topology.

\begin{figure}[!htbp]

  \begin{center}
      \includegraphics[width=0.5\linewidth]{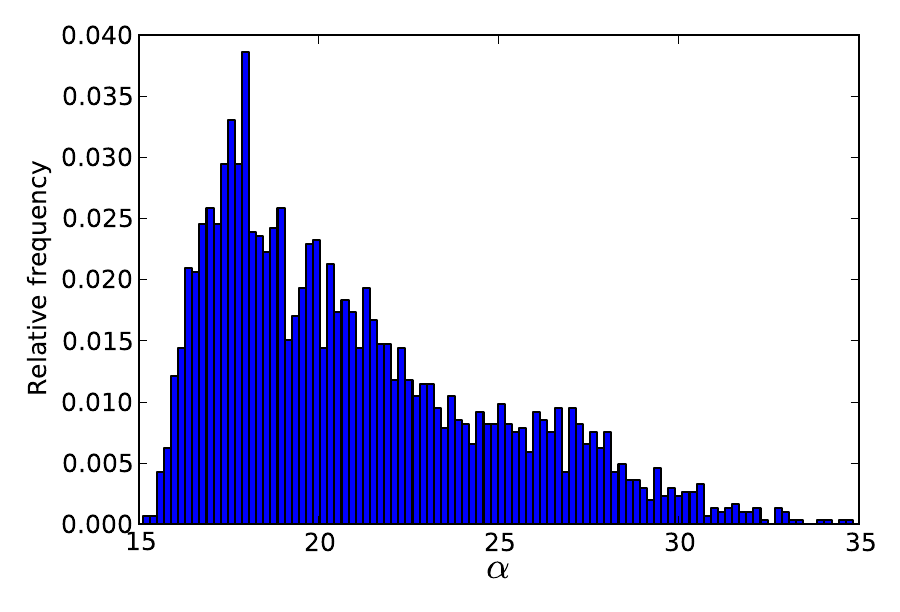}
      \caption{Histogram of $\alpha$ values related to the SIS epidemic model taking place on the airpot network. The first infection time of the nodes was taken as a dynamical feature to calculate $\alpha$.}~\label{f:alfa_epidemia}
  \end{center}
\end{figure}

\begin{figure}[!htbp]

  \begin{center}
      \includegraphics[width=0.85\linewidth]{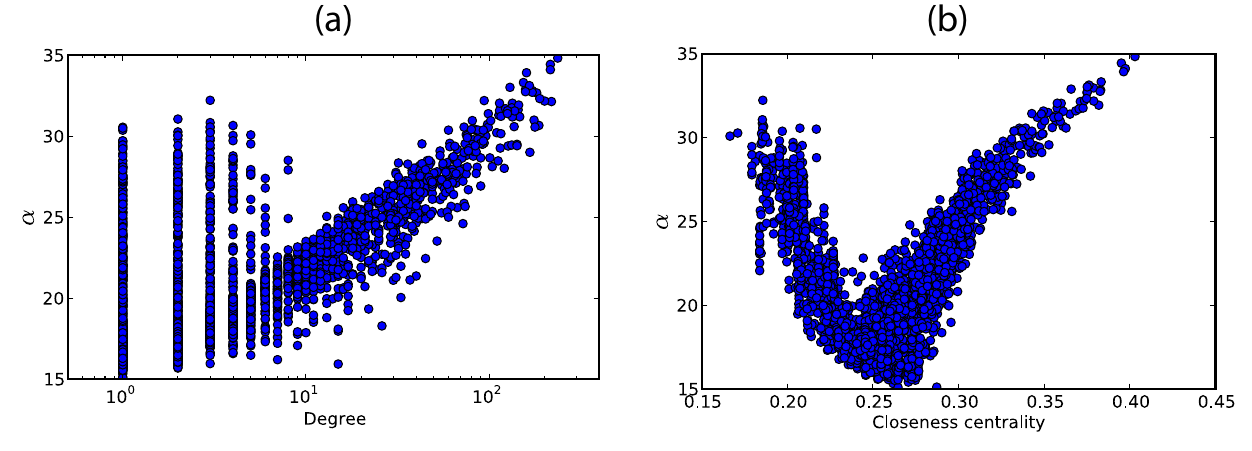}
      \caption{Scatter plots between $\alpha$ and topological features of the nodes. The two features used were (a) the degree and (b) the closenes centrality.}~\label{f:alfa_medidas}
  \end{center}
\end{figure}

We did 1000 simulations of the SIS epidemics dynamics and calculated the respective $\alpha$ values, the result is shown in Figure \ref{f:alfa_epidemia}. The first noticeable feature is the small number of nodes having a large value of $\alpha$, these nodes are probably in regions of the network where the epidemics is far from its average behavior. They could be isolated from the network, and so would have a very small chance of receiving the disease, or they could be hubs of the network, in a way that not mattering where the disease started, they always catch it immediately. We found that the node types are actually mixed, with a little advantage to the hubs. In Figure \ref{f:alfa_medidas}(a) we plot $\alpha$ versus the degree of the nodes, where we see that a node having high degree is guaranteed to have large $\alpha$, while low degree nodes can have a range of $\alpha$ values. This means that high degree nodes depend only on their immediate neighborhood, i.e., the local topology is completely defining their dynamics. On the other hand, low degree nodes are being strongly influenced by their higher order neighborhood, the local topology is not sufficient to define the dynamics. In order to show this more precisely, we plot in Figure \ref{f:alfa_medidas}(b) the relationship of $\alpha$ with the closeness centrality of the nodes \cite{newman:2010}. This topological measure takes into account all network nodes on its calculation, quantifying how central the node is. We see that the closeness defines much better the dynamics of the less connected nodes, showing that they are being differentiated by the topology, but the difference is being caused by the influence of the entire network.

\section{Conclusions}

In a dynamical system underlain by a completely regular topology
(e.g. a toroidal lattice), every node behaves identically regarding its influence on the overall dynamics. It remains an important question to quantify how local heterogeneities in the topology, which can be understood as structural symmetry breaks, may influence the unfolding of the respective dynamics. Despite continuing interest in this area, relatively incipient results have been obtained as a consequence of the fact that several elements that can interfere with the overall dynamics --- such as initial conditions, stochasticity, and parameter configurations --- are not kept constant while inferring individual effects. The current article has addressed this problem by proposing a framework for quantifying to which an extent the topology around each node contributes to differentiating the dynamics. Moreover, the method is devised in such a way as to not require the consideration of any specific topological or structural metric.  Though the method can be applied with respect to any of the potentially interfering elements, in the present work we restrict our attention to the effect of initial conditions.

We illustrated the potential of the reported methodology with respect to a range of different topologies, namely the ER, BA and geographic random models, the neuronal connections of the nematode (\emph{C. elegans}) and the world-wide airport network. Several interesting findings are reported, including the fact that the nodes in ER networks are not significantly differentiated regarding their respective time series. The geographic model exhibits groups of nodes that are highly differentiated in comparison to the majority of network nodes, a consequence of the high statistical fluctuations present in the network construction. The result for the BA model showed that the topological particularities of the hubs are amplified in the dynamics taking place on the system.

Regarding the \emph{C. elegans} network, we found that some nodes
are highly differentiated by their spiking rate. While all these nodes have been found to be well connected nodes, there are well connected nodes that are not in this group, indicating the presence of additional topological influences besides the node degree. We identified these highly active nodes as corresponding to interneurons of the ventral cord of the nematode. For the epidemic dynamics we found that nodes having high degree are isolated from the network influence, in the sense that their dynamics show little variation with different initial conditions. On the other hand, in order to predict the dynamics of low degree nodes, we need information that it is not available on their local neighborhood.

Several future developments are possible, including the consideration of other types of dynamics, other models of networks, as well as investigating the effect of stochasticity and varying parameters or dynamics at each node.

\section*{Acknowledgments}

Luciano da F. Costa is grateful to FAPESP (05/00587- 5) and CNPq (301303/06-1 and 573583/2008-0) for the financial support. C.H.C. is grateful to FAPESP for sponsorship (2011/22639-8). M.P. Viana thanks to FAPESP for financial support (2010/16310-0).

\section*{References}
\bibliographystyle{unsrt}
\bibliography{references_new_version}

\newpage

\section*{Supplementary material of \emph{the relationship between structure and function in complex networks observed locally}}

\renewcommand\thefigure{S\arabic{figure}} 
\setcounter{figure}{0} 

\begin{figure}[!htbp]

  \begin{center}
      \includegraphics[width=0.8\linewidth]{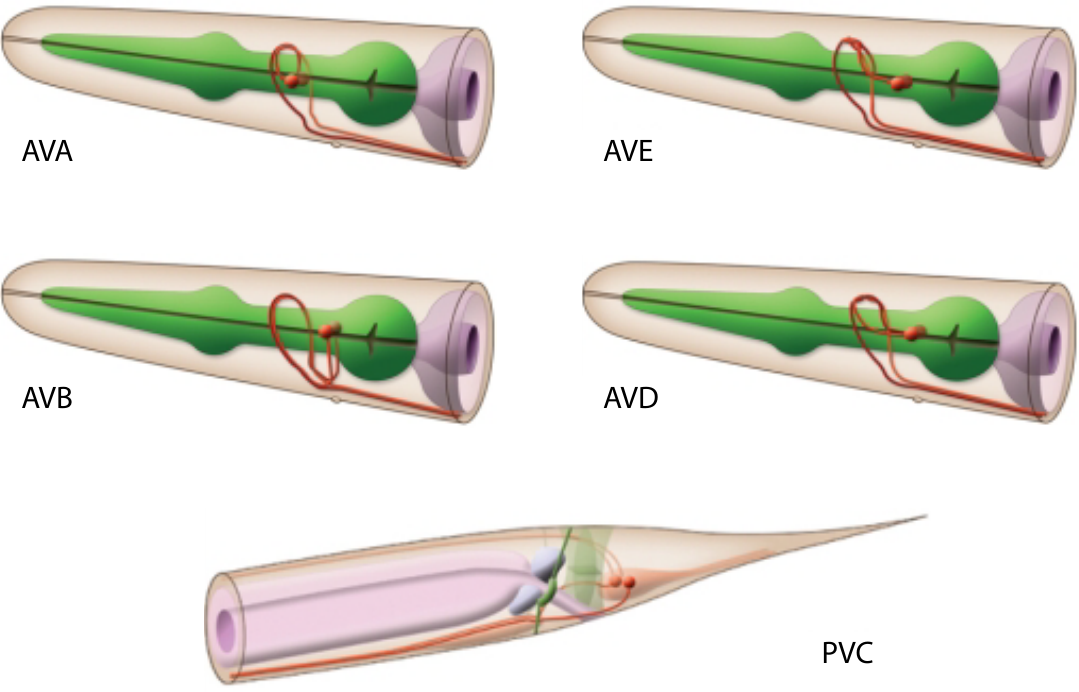}
      \caption{Illustration of the neurons of the \emph{C. elegans} with high dynamical differentiation, i.e., high values of $\alpha$. Dark or light red indicate the left and right version of the neuron. Pictures obtained from \cite{base_elegans}.}~\label{f:neurons}
  \end{center}
\end{figure}

\end{document}